# Approaching the quantum critical point in a highly-correlated all-in-all-out antiferromagnet

Yishu Wang[1,2], T. F. Rosenbaum[1], D. Prabhakaran[3], A. T. Boothroyd[3], Yejun Feng[1,4,*]

[1]Division of Physics, Mathematics, and Astronomy, California Institute of Technology, Pasadena, California 91125, USA
[2]The Institute for Quantum Matter and Department of Physics and Astronomy, The Johns Hopkins University, Baltimore, Maryland 21218, USA
[3]Department of Physics, University of Oxford, Clarendon Laboratory, Oxford, OX1 3PU, United Kingdom
[4]Okinawa Institute of Science and Technology Graduate University, Onna, Okinawa 904-0495, Japan
*Correspondence author. Email: <yejun@oist.jp>.

**ABSTRACT:**
**Continuous quantum phase transitions involving all-in-all-out (AIAO) antiferromagnetic order in strongly spin-orbit-coupled 5$d$ compounds could give rise to various exotic electronic phases and strongly-coupled quantum critical phenomena. Here we experimentally trace the AIAO spin order in $Sm_2Ir_2O_7$ using direct resonant x-ray magnetic diffraction techniques under high pressure. The magnetic order is suppressed at a critical pressure $P_c$=6.30 GPa, while the lattice symmetry remains in the cubic $Fd$-$3m$ space group across the quantum critical point. Comparing pressure tuning and the chemical series $R_2Ir_2O_7$ reveals that the suppression of the AIAO order and the approach to the spin-disordered state is characterized by contrasting evolutions of both the pyrochlore lattice constant $a$ and the trigonal distortion $x$. The former affects the 5$d$ bandwidth, the latter the Ising anisotropy, and as such we posit that the opposite effects of pressure and chemical tuning lead to spin fluctuations with different Ising and Heisenberg character in the quantum critical region. Finally, the observed low-pressure scale of the AIAO quantum phase transition in $Sm_2Ir_2O_7$ identifies a circumscribed region of $P$-$T$ space for investigating the putative magnetic Weyl-semimetal state.**

The mix of magnetic interactions, electron correlations, and spin-orbit coupling informs the competition between different quantum ground states and ordering mechanisms, ranging from Mott to Slater antiferromagnetic insulators [1-2], phonon- to spin-fluctuation-mediated superconductivity [3-4], and Kondo screening to RKKY exchange in heavy fermion materials [5]. For 5$d$ pyrochlores such as $R_2Ir_2O_7$ ($R$=Y, Eu, Sm, Nd), the interplay between intermediate electron correlations and strong spin-orbit coupling leads to all-in-all-out (AIAO) antiferromagnetic order and, potentially, non-trivial topological band structure, commonly known as a Weyl semimetal of broken time reversal symmetry [6-10]. Conversely, without electron correlation, strong spin-orbital coupling could induce a different topological Weyl state of broken inversion symmetry, as proposed in non-magnetic pyrochlores with a breathing lattice [11].

AIAO spin order in 5$d$ pyrochlores has been verified experimentally in both $R_2Ir_2O_7$ ($R$=Lu, Yb, Tb, Eu, Sm, Nd) and $Cd_2Os_2O_7$ [12-17]. However, recent angle-resolved photoemission measurements on both magnetic $Nd_2Ir_2O_7$ and nonmagnetic $Pr_2Ir_2O_7$ as the end member of the $R_2Ir_2O_7$ series demonstrate parabolic nodal structures [2, 18] that raise questions about the existence of magnetic Weyl semimetal phases in these compounds at ambient pressure. Under suitable tuning processes, such as pressure, exotic states may yet emerge over an adjustable parameter space spanned by the Coulomb interaction $U$ and spin-orbit coupling $\lambda$, normalized to the hopping integral $t$ [18, 19]. Given that the presence of AIAO magnetic order serves as a gauge of electron correlations, its quantum critical point, where the magnetic order is suppressed to zero at zero temperature, could identify some of the most intriguing regions of intermediate to strong-coupling physics in $5d$ compounds [6-10]. For example, it has been suggested that $Pr_2Ir_2O_7$ develops a two-in-two-out spin ice configuration that melts into a metallic spin liquid at $T < 0.4$ K in the proximity of its AIAO quantum critical point [20].

The iridate pyrochlores $R_2Ir_2O_7$ provide a series of model systems susceptible to continuous pressure tuning, with an approximately local $J_{eff}=1/2$ moment from the $Ir^{4+}$ ions of the $5d$ $t_{2g}$ band [15] and a number of germane theoretical calculations [6-10]. $Sm_2Ir_2O_7$, with proven AIAO order [14], negative pressure dependence of its insulating phase [21], and available high-quality single crystals, is a particularly promising experimental choice. As we demonstrate below, the AIAO spin order in $Sm_2Ir_2O_7$ experiences a continuous quantum phase transition at a modest critical pressure $P_c=6.30$ GPa in the presence of constant lattice symmetry, exemplifying the first directly-tracked AIAO quantum critical point in iridates under pressure. Moreover, the pressure evolution of $Sm_2Ir_2O_7$ follows a different pathway across the $U/t$-$\lambda/t$ phase space compared with the $R_2Ir_2O_7$ chemical series, providing further clues to the nature of intermediate to strong-coupling physics in this model system.

We directly probe the AIAO spin order in $Sm_2Ir_2O_7$ under pressure using resonant x-ray magnetic diffraction at beamline 4-ID-D of the Advanced Photon Source ([22-23] and Supplementary materials). The pyrochlore structure in the $Fd$-$3m$ space group is fully characterized by two parameters, the lattice constant $a$ and the coordination parameter $x$ [9]. From the single-peaked (1, 1, 1) and (2, 2, 0) diffraction orders, $Sm_2Ir_2O_7$ remains in a cubic structure to at least 21 GPa, and the lattice constant $a(P)$ evolves continuously at 4 K without any visible sign of a phase transition (Fig. 1). The simple lattice evolution strongly suggests a continuous AIAO quantum phase transition, motivating further polarization-analysis of resonantly scattered x-ray magnetic diffraction signals (Figs. 2-4). The cubic space group under pressure is illustrated by measured (0, 0, 6) and (0, -2, 4) diffraction intensities in the $\pi$–$\pi$' channel, which are minimal and constant through 21 GPa (Fig. 4b). Thus the $Fd$-$3m$ space group persists at 4 K, ruling out a breathing lattice instability of the $F$-$43m$ space group type as observed in $Cd_2Os_2O_7$ [24].

Within the $Fd$-$3m$ space group, both Sm and Ir ions in $Sm_2Ir_2O_7$ do not contribute to diffraction intensities of (1, 1, 1) and (2, 2, 0) orders. As these diffraction intensities arise solely from oxygen ions, the parameter $x$ can be measured with high sensitivity [24]. For single crystals under high pressure, where a full structure refinement is not practical due to time and geometry constraints, measurements of these peaks are especially suitable to reveal the pressure evolution of $x$ [24]. The normalized diffraction intensities in the $\pi$–$\pi$' channel appear constant under pressure (Figs. 1b, 1c), implying a stable $x$ over 21 GPa in $Sm_2Ir_2O_7$.

The ATS resonance in the $\pi$–$\sigma$ channel, measured at the (2, 4, 0) order and $\psi$~0°, demonstrates a constant shape under pressure. In $R_2Ir_2O_7$, the ATS resonance profile differs in shape from that of the magnetic resonance (Fig. 3 and Ref. [13-14]), indicating that the magnetic electrons are confined in the lower $t_{2g}$ band. The ATS resonance is sensitive to the individual $t_{2g}$ and $e_g$ bands of Ir $5d$ states, and our result in Fig. 1d demonstrates that both bands experience no significant energy shift over this pressure range. The constant behavior of ATS resonance is also observed in $Cd_2Os_2O_7$ under pressure [24].

We show in Fig. 2 raw magnetic diffraction profiles in the π–σ channel of both the sample mosaic and energy resonance of the (0, 0, 6) order at pressures across the magnetic quantum phase transition. In addition to the evolution with $P$ at fixed $T = 4$ K, we explore the temperature evolution of the (0, 0, 6) order in $Sm_2Ir_2O_7$ at $P = 6.26$ GPa, just below $P_c$. The mosaic profile is measured up to 40 K for one azimuthal condition, with the energy resonance profile measured at selected $T$ (Fig. 3a). The integrated diffraction intensity continuously approaches a constant beyond 27 K (Fig. 3b), demonstrating a second-order thermal AIAO phase transition at 6.26 GPa.

High-resolution longitudinal scans of the (0, 0, 6) order at 6.26 GPa and both 4 and 23 K indicate that $Sm_2Ir_2O_7$ still has long-range AIAO order, as the diffraction line shapes are instrument resolution-limited with a spin coherence length of at least 1450 Å. At 27 K, the line shape broadens to a diffusive shape, indicating a shortened spin correlation length of approximately 450 Å at the magnetic transition. Furthermore, our high-resolution study of the 2θ value of the (0, 0, 6) diffraction reveals that the lattice constant $a$ shrinks with increasing $T$ from 4 to 27 K due to a decreasing $<M>$ (Fig. 3b inset). This anomalous $a(T)$ is a reflection of the overall magnetostriction, which was also observed in the antiferromagnets $Nd_2Ir_2O_7$ and $NiS_2$ at ambient pressure [25, 26]. As $a(T)$ evolves similarly to the diffraction intensity in Fig. 3b, there is a consistent relationship of $\Delta a(T) \sim I_{(006)}(T) \sim <M>^2$. The magnetostriction $\Delta a/a \sim 5 \cdot 10^{-4}$ in $Sm_2Ir_2O_7$ ($T_N$=26.7K) is comparable in size to that in $Nd_2Ir_2O_7$ ($T_N$=33K) [25]. At $P$=6.7 GPa, above $P_c$, there is no observed magnetic resonance (Fig. 2d), and a similar temperature study of the mosaic profile generates no temperature dependence up to 20 K.

We fit both the thermal and pressure evolution (Figs. 3b, 4a) of the resonant magnetic diffraction intensity to critical power law forms: $I_{(006)} \sim (T_c-T)^{2\beta}$ and $I_{(006)} \sim (P_c-P)^{2\gamma}$. We find $T_c$=26.8±0.3 K and $\beta$=0.41±0.05 at $P$=6.26 GPa, and $P_c$=6.30±0.05 GPa and $\gamma$=0.15±0.03 at $T$=4.0 K. The exponent $\beta$ is between the mean-field expectation of 0.5 and three-dimensional Heisenberg spin fluctuations of 0.37, but the order parameter evolves more rapidly under pressure with a small $\gamma$.

For the $P$-$T$ phase diagram of AIAO order in $Sm_2Ir_2O_7$, we scale $T_N(P)$ by the magnetic diffraction intensity $I_{(006)}$ in Fig. 4a. This mapping of magnetic intensity to the phase boundary is justified by the consideration that within this small pressure range ($\Delta a \sim 0.10$Å or $\Delta a/a \sim 1.0$%), the order parameter strength, defined as the staggered moment $<M>$, should connect to the energy scale of $T_N$ as $I_{(006)}(P) \sim <M>^2 \sim T_N(P)$. This relation has been demonstrated previously in several antiferromagnets under pressure [4, 24]. The projected phase boundary $T_N(P)$ is consistent with the three observed phase points (Fig. 4c), identified through magnetization $M(T)$ at ambient $P$, the temperature dependence of $I_{(006)}(T)$ at $P$=6.26 GPa, and the pressure dependence of $I_{(006)}(P)$ at $T$=4 K. At ambient pressure, Sm moments in $Sm_2Ir_2O_7$ have an estimated size of 0.1 $\mu_B/Sm^{3+}$, and order at $T \sim 10$K [27]. Both are much smaller than the Ir moment size of 0.3 $\mu_B/Ir^{4+}$ and the ordering temperature $T_N \sim 110$ K [27]. By comparison to several pyrochlore iridates with large $A$-site moments (2.6-9 $\mu_B$ per $Nd^{3+}$, $Er^{3+}$, or $Tb^{3+}$) [28, 29], the magnetic coupling strength between $Sm^{3+}$ $4f$ moments is likely much below 0.1 meV [28], and $Sm^{3+}$ ordering would rely on the assistance of the $Ir^{4+}$ molecular field, making it parasitic to the Ir AIAO order. We thus consider $Sm^{3+}$ ions as disordered at $P_c$=6.30 GPa and $T$=4 K.

Although both pressure and chemical variation of $R$ (from Eu, Sm, Nd, to Pr) in $R_2Ir_2O_7$ [19] are effective in suppressing $T_{N/MIT}$ to zero while the lattice persists in the cubic $Fd$-$3m$ symmetry, there exist microscopic differences between these two tuning mechanisms. For the two structural parameters $a$ and $x$ of the pyrochlore lattice [9], $a$ decreases ~ 1.0% at $P_c$ in $Sm_2Ir_2O_7$, but increases ~ 1.5% in the chemical series from Eu to Pr (Fig. 4d). The parameter $x$ indicates compressive trigonal distortion of the octahedron $IrO_6$. For $Eu_2Ir_2O_7$ and $Pr_2Ir_2O_7$ at ambient pressure, $x$ reduces from 0.339 to 0.330 with chemical variation (Fig. 4d). However, $x$ is largely constant if not increasing in $Sm_2Ir_2O_7$ under pressure (Fig. 1). In other $5d$ AIAO pyrochlores, $x$ increases from 0.330 to 0.335 in $Eu_2Ir_2O_7$ over 17 GPa at 295 K [30], and from 0.319 to about 0.325 in $Cd_2Os_2O_7$ over 40 GPa at 4 K [24].

AIAO order exists in pyrochlore lattices due to both the electron correlation, $U/t$, and local Ising spin anisotropy. From symmetry considerations, the pyrochlore structure naturally hosts the Dzyaloshinskii-Moriya interaction, and the *direct* DM interaction leads to the AIAO order [31]. As the DM strength is proportional to the spin-orbit coupling $\lambda$, AIAO order has been discovered in many $5d$ pyrochlores. The trigonal distortion of local crystal fields is of similar strength to the spin-orbit coupling [14, 32], and varies the Ir-O-Ir bond angle to affect all ranks of the super-exchange interaction. The increasing $x$ sharpens the Ir-O-Ir bond angle from ~131.3° at $x$=0.330 in $Pr_2Ir_2O_7$ to ~126.7° at $x$=0.339 in $Eu_2Ir_2O_7$. By contrast, a shrinking $a(P)$ directly increases the $5d$ bandwidth or equivalently the hopping strength $t$. A 3% volume reduction at 6.7 GPa would inject ~600 meV energy into each unit cell [22], presumably distributed among all valence electrons at the Fermi surface, with half to broaden the Ir $5d$ band. The increased spatial extent of Ir $5d$ orbitals under pressure reduces $U/t$. Conversely, the lattice expansion in the chemical series $R_2Ir_2O_7$ reduces the hopping strength $t$ towards the paramagnetic metal. We thus expect $t$ to be predominantly affected by the lattice constant, while the spin anisotropy is controlled by $x$.

The contrasting effects of pressure and chemical tuning now become clear and are captured in Fig. 4e. Pressure maintains the axial nature of the local spin anisotropy, but increases the hopping integral $t$ and reduces $U$ to suppress the long-range order. Chemical tuning of $R_2Ir_2O_7$ suppresses the AIAO state by reducing the axial spin anisotropy towards a more isotropic Heisenberg spin state, mainly through a weakened Dzyaloshinskii-Moriya interaction from a more obtuse Ir-O-Ir angle [7, 10]. There are likely two separate pathways across the quantum phase boundary in $U/t$-$\lambda/t$ parameter space (Fig. 4e) [19], accompanied by different types of spin fluctuations at the respective critical points.

A metal-insulator transition runs concurrently with the AIAO magnetic order in $R_2Ir_2O_7$ ($R$= Eu, Sm, and Nd) at ambient pressure, and was measured in pelleted polycrystalline $Sm_2Ir_2O_7$ up to 2.2 GPa [21]. The pressure suppression of the insulating phase is consistent with our measured AIAO magnetic phase boundary (Fig. 4c) for the limited region they overlap. Several theoretical simulations [6-7, 9] have suggested the existence of a Weyl semimetallic AIAO phase between the AIAO Mott insulator and the paramagnetic metal. While the Coulomb interaction $U$ varies from 0.5 to 2 eV in various models, there seems to be agreement that the Weyl semimetal phase spans a finite width of $\Delta U \sim 0.2$ eV. As the pressure-driven AIAO quantum phase transition happens within ~0.3 eV change in $t$ from the ambient condition, a span of $\Delta U \sim 0.2$ eV would likely cover the whole pressure range of AIAO order evolution in $Sm_2Ir_2O_7$. Nevertheless, given the small critical exponent $\gamma$ and the strongly convex shape of the $P$-$T$ phase diagram, the electronic structure might well only demonstrate topological features very close to the pressure phase boundary (if at all).

$Sm_2Ir_2O_7$ represents one of the cleanest systems to explore the AIAO type of antiferromagnetic quantum criticality with Ir moments maintaining local Ising anisotropy under pressure and being much larger in size than the Sm moments. With no breaking of inversion symmetry through the quantum critical point, it provides a fascinating comparison to the strongly-coupled AIAO quantum phase transition in $Cd_2Os_2O_7$ [24]. Although the electronic evolution through $P_c$ remains to be resolved, most theoretical simulations agree in general that a metallic paramagnetic phase exists beyond the AIAO order. Whether or not there exists a magnetic Weyl semimetal phase, a strong electronic evolution likely exists close to the AIAO quantum critical point, providing the quantum critical region strong-coupling characteristics with intertwined spin and charge fluctuations.

**Acknowledgments**
Y.F. acknowledges the support from Okinawa Institute of Science and Technology Graduate University with subsidy funding from the Cabinet Office, Government of Japan. The work at Caltech was supported by National Science Foundation Grant No. DMR-1606858. The work in Oxford was supported by UK Engineering and Physical Sciences Research Council grant no. EP/N034872/1. The work at the Advanced Photon Source of Argonne National Laboratory was supported by the US Department of Energy Basic Energy Sciences under Contract No. NEAC02-06CH11357.

**Figure captions:**

Fig. 1. (a) Lattice constant $a(P)$ at 4.0±0.3 K is fit to a two-parameter Birch equation, with a bulk modulus $B_0$=215.6±4.8 GPa, and its derivative $B'$=3.9±0.5. (inset) Representative longitudinal (θ-2θ) scans of the (2, 2, 0) and (1, 1, 1) orders at five pressures, showing single peaks with minimal traces of a stressed lattice condition. Our measured $B'$ is much smaller than that of the silver manometer, in sharp contrast to several accounts of large $B'$ values of iridate pyrochlores in the literature [30]. (b-c) Diffraction intensities of the (2, 2, 0) and (1, 1, 1) orders, normalized by those of the (4, 4, 0) and (2, 2, 2) orders, respectively. Different symbols (circle, square, and diamond) in Figs. 1b, 1c, 4a, and 4a represent each of three individual samples studied under pressure. (d) The ATS resonance at four pressures and 4 K. The spectral shape has no azimuthal dependence, from measurements at (2, 4, 0) and (0, -2, 4) orders with ψ~0° and ~35°, respectively, up to 21 GPa (not shown).

Fig. 2. Raw x-ray magnetic diffraction profiles of both mosaic and energy resonance at both (a-d) below, and (e-f) above the critical pressure $P_c$ = 6.30 GPa from one sample. The common spectral weight (marked by solid red circles) between energy spectra at all azimuthal angles, defines the multiple-scattering-free resonance profile. All counting rates are normalized to a 100-mA synchrotron ring-current in Figs. 2 and 3.

Fig. 3. Temperature evolution of the resonant x-ray magnetic diffraction at 6.26 GPa, measured at the (0, 0, 6) order in the π–σ channel. (a) Mosaic scans and energy profiles. Above 27 K, the magnetic resonance fully disappears and the mosaic profile no longer varies. Given the similar shapes of three mosaic profiles at azimuthal ψ=137°~140° in Fig. 2c, the residual mosaic form is likely due to dislocations and voids, instead of multiple scattering, and could be attributed to a small σ component in the π–polarized synchrotron light, sharing the same origin as the minimal (0, 0, 6) diffraction intensity in the π–π' channel (Fig. 4b). (b) Integrated mosaic intensity in Fig. 3a as a function of temperature. The data is fit to a power law plus a constant. (inset) Lattice expansion at low temperature demonstrates a noticeable magnetostriction effect. (c) Longitudinal scans of the (0, 0, 6) order at three different temperatures, fit to resolution-limited Pseudo-Voigt line shapes at 4 K and 23 K, and a diffusive Lorentzian shape plus a linear background at 27 K.

Fig. 4. (a) Integrated magnetic diffraction intensity of the (0,0,6) order in the π–σ channel, normalized by that of the (0,0,4) order in the π–π' channel. A fit to a power law plus a constant (solid line) reveals the AIAO quantum phase transition at $P_c$=6.30 GPa at $T$=4.0±0.3 K. (b) Integrated intensities of the (0,-2,4) and (0,0,6) orders in the π–π' channel, normalized by that of the (0,0,4) order, indicate the *Fd-3m* space group persists to 21 GPa. (c) The projected *P-T* phase diagram of $Sm_2Ir_2O_7$ (gray area) is scaled from the power-law fit of the magnetic intensity in (a), with magnetic phase boundary points measured on our samples (orange symbols), and metal-insulator transition points (green circles) from the literature [23]. (d) Correlation between pyrochlore lattice parameters $x$ and $a$ in $R_2Ir_2O_7$ series for elements $R$ = Gd (down triangle), Eu (diamonds), Sm (squares), Nd (circles), and Pr (up triangles), from seven independent research groups (Supplementary Table I, Ref. [21, 25, 30, 34-37]). ($x$, $a$) values of the same group but of different element $R$ are connected by linear segments, highlighting the correlation between $x$ and $a$ (grey dashed line) despite systematic variations between different crystal growers. (e) Evolution of the AIAO order in the three-dimensional *T-a-x* phase space as a function of $P$ and $R$, including $Sm_2Ir_2O_7$ under pressure (red solid circles) and the $R$ series (solid blue squares) with the generally agreed $T_N$ [19], the average lattice constant for each $R$ element, and the grey *x-a* line in (d). $T_N$ of doping series (Sm, Nd, Pr)$_2$Ir$_2$O$_7$

(blue open squares), and pressure series of $Nd_2Ir_2O_7$ (orange open circles) are assumed from the metal-insulator transitions in Ref. [38]. The striking differences of curvature, convex vs concave, at the different quantum phase transitions points to the different roles played by correlations under doping and pressure.

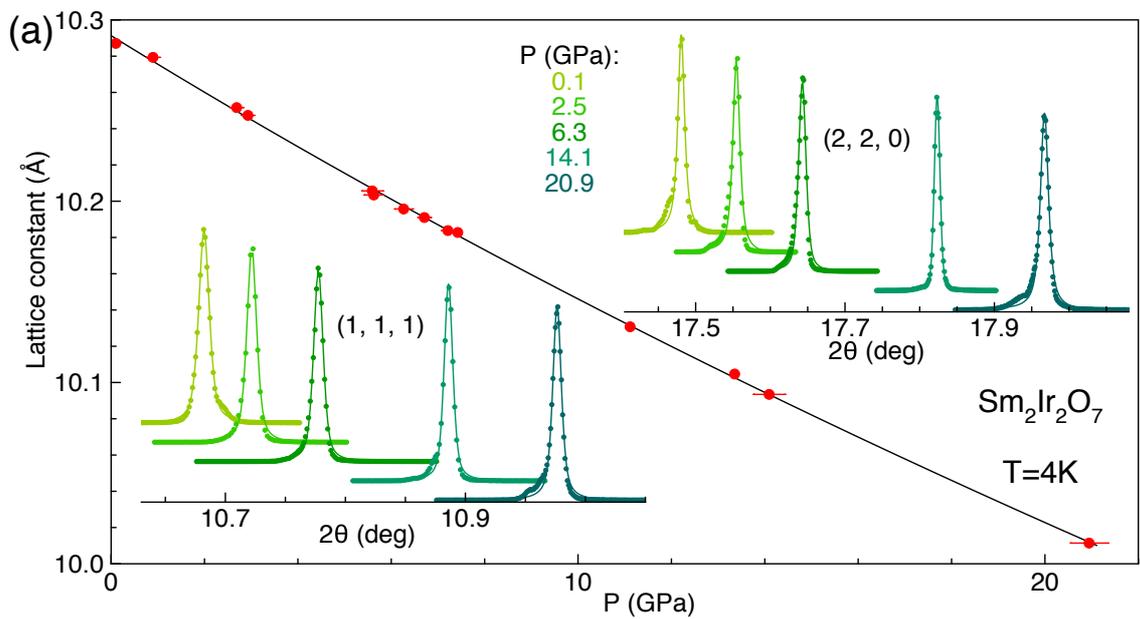
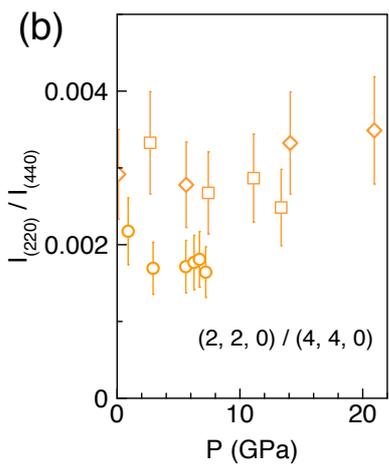
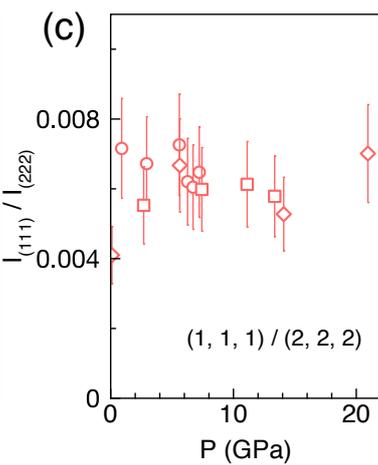
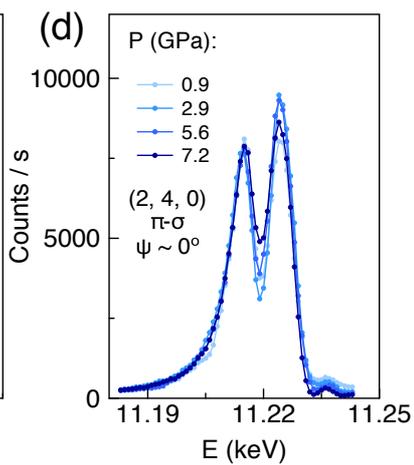

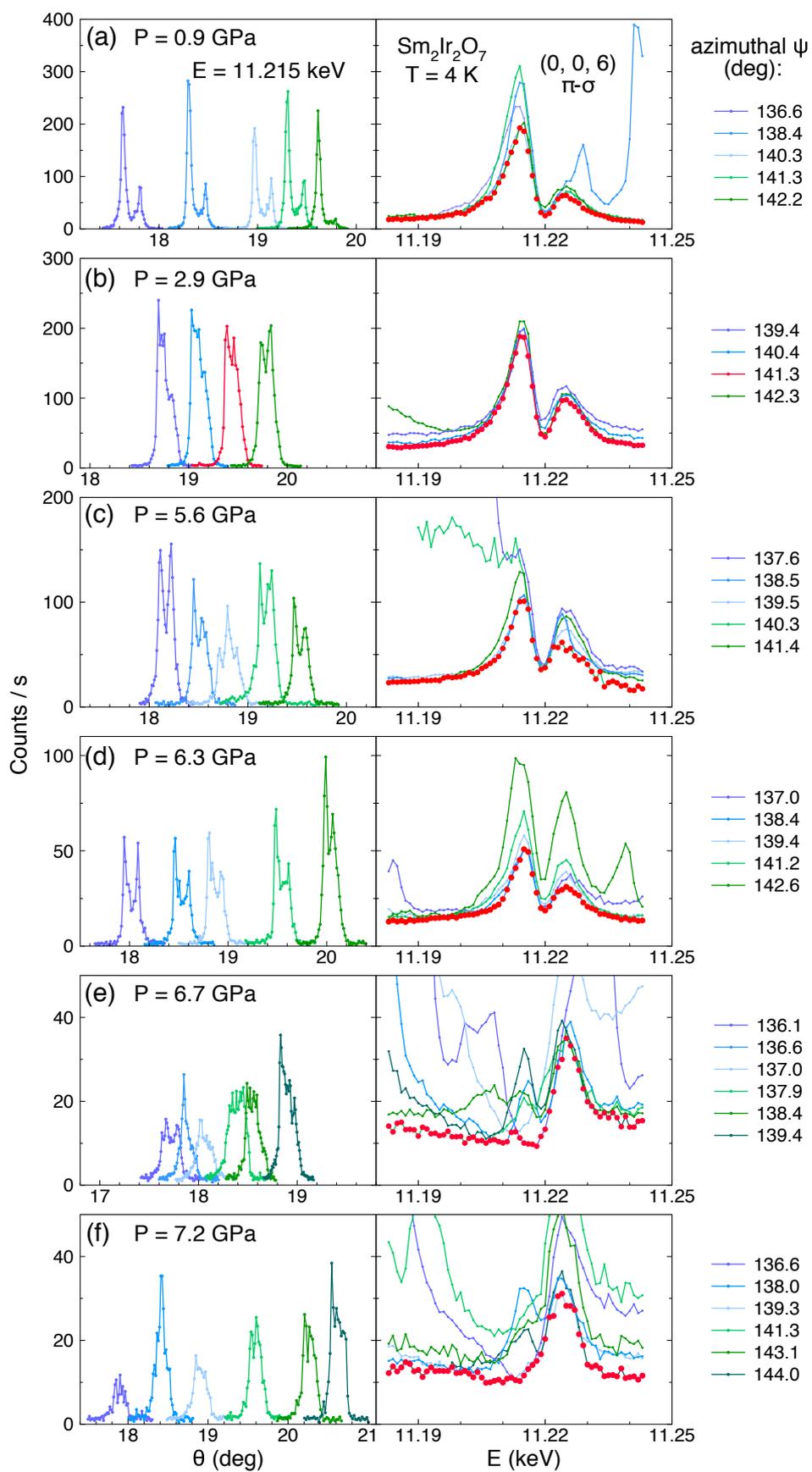

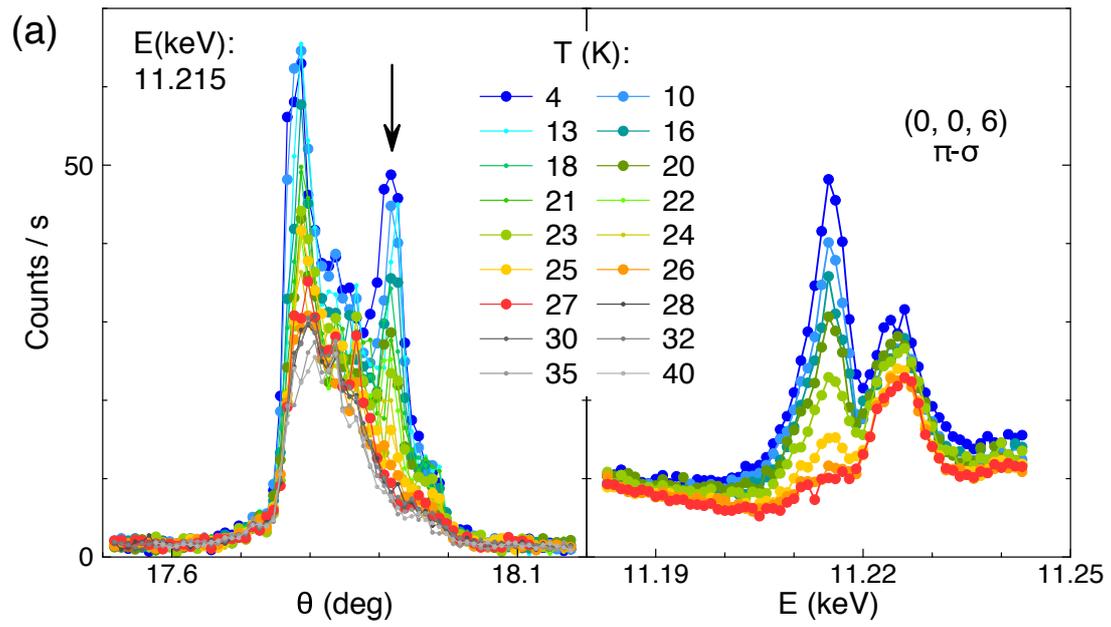

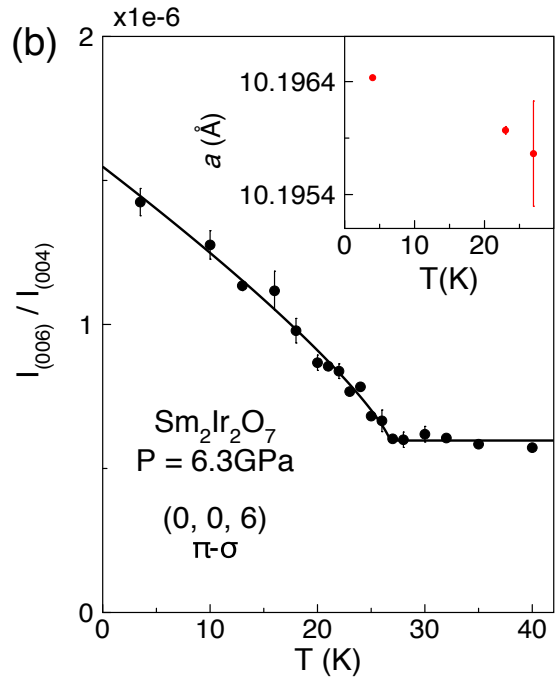

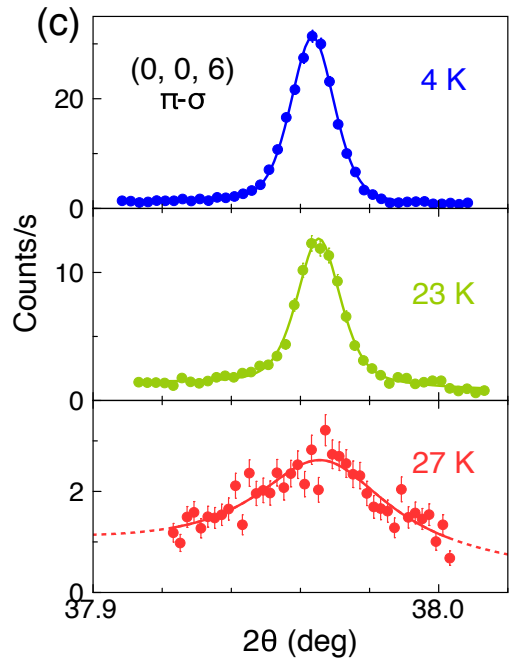

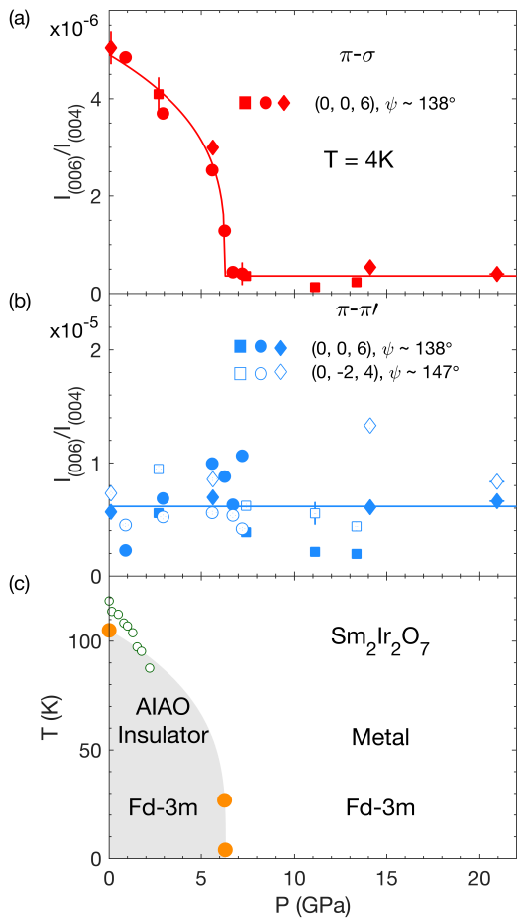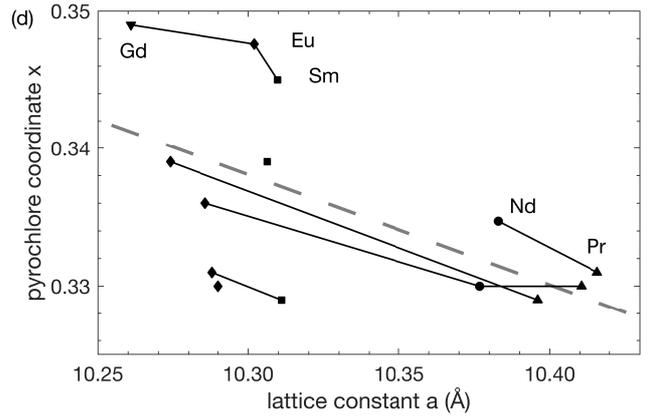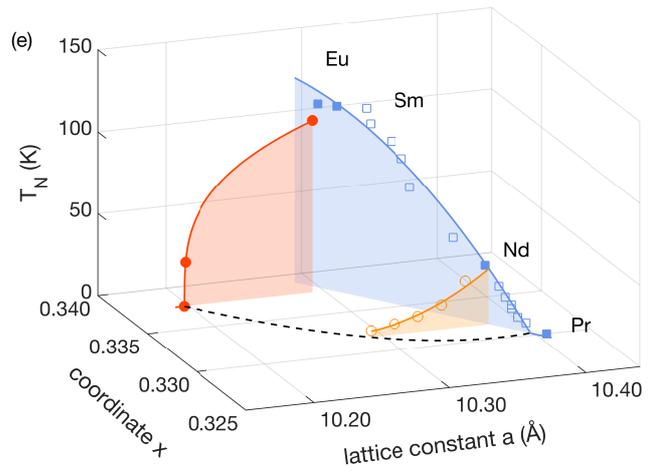

**Supplementary materials:**

**Approaching the quantum critical point in a highly-correlated all-in-all-out antiferromagnet**

Yishu Wang[1,2], T. F. Rosenbaum[1], D. Prabhakaran[3], A. T. Boothroyd[3], Yejun Feng[1,4,*]

[1]Division of Physics, Mathematics, and Astronomy, California Institute of Technology, Pasadena, California 91125, USA
[2]The Institute for Quantum Matter and Department of Physics and Astronomy, The Johns Hopkins University, Baltimore, Maryland 21218, USA
[3]Department of Physics, University of Oxford, Clarendon Laboratory, Oxford, OX1 3PU, United Kingdom
[4]Okinawa Institute of Science and Technology Graduate University, Onna, Okinawa 904-0495, Japan

**Experimental Methods:**

We directly probe the AIAO order in $Sm_2Ir_2O_7$ under pressure at beamline 4-ID-D of the Advanced Photon Source [22-23]. *L*-edge resonance in Ir compounds is limited to the $L_3$ edge ($E$=11.215 keV) due to the presence of strong spin-orbit coupling. Using a transmission (Laue) geometry in the horizontal plane, and plate-shaped single crystal samples of 15-20 μm thickness with a surface normal of (1, -1, 0), we could access diffraction orders such as (0, 0, 6), (1, 1, 1), (2, 2, 0), (0, -2, 4), and (2, 4, 0) within the confined geometry of a diamond anvil cell. The charge resonance, known as the anisotropic tensor susceptibility (ATS), has an energy profile with contributions from both $5d$ $t_{2g}$ and $e_g$ bands across the $L_3$ edge (Fig. 1d) [13, 14]. The spin resonance, with a contribution only from the $5d$ $t_{2g}$ band, has a different profile mostly confined below the $L_3$ edge (Fig. 2). While both spin and charge resonances are observed at forbidden lattice diffraction orders such as (0, 2, 4) and (0, 0, 6), the magnetic resonance signal can be isolated from the charge resonance at (0, 0, 4n+2) orders with a special azimuthal ψ~45° relative to the (1, 0, 0) vector [12-14, 22, 24].

Our general high-pressure and x-ray techniques have been discussed in Refs. [22-24]. A methanol/ethanol 4:1 mixture was used as the pressure medium. Pressure was calibrated by a Ag manometer *in situ* at $T$ = 4 K using a two-parameter Birch equation of state, with bulk modulus $B_0$=108.85 GPa and its derivative $B'$=$dB$/$dP$=5.7 over a 20 GPa range. X-ray polarization analysis was provided by a highly oriented pyrolytic graphite crystal of 0.35° FWHM mosaic at (0, 0, 8) order, in order to match up to the typical sample mosaic under pressure. Data presented here were collected from three $Sm_2Ir_2O_7$ samples, prepared from three different octahedral shaped single crystals in the same growth batch, and all pressurized across the AIAO magnetic phase boundary with both spin and charge resonances explored.

**Supplementary Table I:**
A compendium of lattice parameters of $R_2Ir_2O_7$ plotted in Fig. 5a.

| Element $R$: | Lattice constant (Å): | Pyrochlore parameter x: | Reference: |
|---|---|---|---|
| Gd | 10.2609 | 0.3490 | 21 |
| Eu | 10.3020 | 0.3476 | 21 |
| Eu | 10.290 | 0.330 | 30 |
| Eu | 10.2857 | 0.336 | 25 |
| Eu | 10.2740 | 0.339 | 34 |
| Eu | 10.2880 | 0.331 | 36 |
| Sm | 10.3097 | 0.3450 | 21 |
| Sm | 10.3063 | 0.339 | 37 |
| Sm | 10.3110 | 0.329 | 36 |
| Nd | 10.3768 | 0.330 | 25 |
| Nd | 10.383 | 0.3347 | 35 |
| Pr | 10.4105 | 0.330 | 25 |
| Pr | 10.3960 | 0.329 | 34 |
| Pr | 10.4159 | 0.3310 | 35 |